\documentclass[10pt,journal,compsoc]{IEEEtran}

\ifCLASSOPTIONcompsoc
\usepackage[nocompress]{cite}
\else
\usepackage{cite}
\fi

\ifCLASSINFOpdf
\else
\fi

\usepackage{graphicx}
\usepackage{listings}
\begin{document}
	
	\title{Revising the classic computing paradigm\\and its technological implementations}
	%
	%
	%
	%
	
	\author{J\'anos V\'egh
		\IEEEcompsocitemizethanks{\IEEEcompsocthanksitem Kalimános BT, Hungary\protect\\
			E-mail: Vegh.Janos@gmail.com}
		\thanks{The paper is an extended and contracted form of presentations held at conferences CSCE and CSCI in Las Vegas, years 2019 and 2020. Manuscript received April 19, 2005; revised August 26, 2015.}}

	\markboth{Trans Computers,~Vol.~14, No.~8, August~2021}%
	{V\~egh \MakeLowercase{}: Revising the classic computing paradigm}

	\IEEEtitleabstractindextext{%
		\begin{abstract}
			Today's computing is told to be based on the classic paradigm, proposed by von Neumann, a three-quarter century ago.
			However, that paradigm was justified (for the timing relations of) vacuum tubes only. The technological development invalidated the classic paradigm (but not the model!) and led to catastrophic performance losses in computing systems, from operating gate level to large networks, including the neuromorphic ones. The paper reviews the critical points of the classic paradigm and scrutinizes the confusion made around it. It discusses some of the consequences of improper technological implementation, from the shared media to the parallelized operation. The model is perfect, but it is applied outside of its range of validity. The paradigm is extended by providing the "procedure" that enables computing science to work with cases where the transfer time is not negligible apart from processing time.
		\end{abstract}

		}

	\maketitle

	\IEEEdisplaynontitleabstractindextext

	\IEEEpeerreviewmaketitle

	\IEEEraisesectionheading{\section{Introduction}\label{sec:introduction}}

	\IEEEPARstart{T}{oday's} 
	computing science commonly refers to
	von Neumann's classic "First Draft"~\cite{EDVACreport1945} as its solid base.
	However, von Neumann performed a general analysis of computing, and --for the intended vacuum tube implementation only-- he made some very strong omissions. He limited the validity of his "\textit{procedure}" (but not of his \textit{model}!) for vacuum tubes only; furthermore,
	made clear that using "too fast" vacuum tubes (or other elementary processing switches) \textit{vitiates} his simplified model. Furthermore, he emphasized that it would be \textit{unsound} to apply his simplified model to modeling neural operation, where the timing relations are entirely different: the conduction [transmission]  time not only not negligible apart from synaptic [processing] time, but is typically much longer than the synaptic [processing] time. In that phase of development, and the age of vacuum tube technology, he did not feel the need to work out the "procedure" that can be followed when the processing elements get faster, and the timing relations do not enable us to neglect transmission time
	apart from processing time. However, he strongly emphasized that the computing paradigm (especially the omissions he made about the timing relations) must be revised according to technology development.
	
	The technology, with the advent of transistors and integrated circuits, quickly forgot the vacuum tubes.
	The stealthy nature of the impressive technological development covered for decades that the computing paradigm, created with vacuum tubes' timing relations in mind, was not valid for the new technology.
	It was early noticed that the development of computing
	has slowed down dramatically~\cite{ComputingPerformanceBook:2011}.
	Many experts suspected that the computing paradigm itself, "\textit{the implicit   hardware/software  contract}~\cite{AsanovicParallelCACM:2009}", is responsible for the experienced issues:
	"\textit{No current programming model is able to cope with this development [of processors], though, as they essentially still follow the classical van Neumann model}"~\cite{SoOS:2010}.
	However, when thinking about "advances beyond 2020", on one side, the solution was expected from the "\textit{more efficient implementation of the von Neumann architecture}"~\cite{DeBenedictis_supercomputing:2014}.
	On the other side, it was stated that "\textit{The von Neumann architecture} is fundamentally inefficient
	and non-scalable for representing massively interconnected neural networks"~\cite{TrueNorth:2016}.
	Anyhow, \textit{it is worth scrutinizing both the classic computing paradigm and its technical implementation}.

	\section{Neumann's ideas} \label{sec:ideas}
	His publication was a source of numerous confusion by his successors.
	
	\subsection{The "von Neumann architecture"}
	The great idea of von Neumann was that he defined an interface (an abstract model) between the physical device and its mathematical exploitation~\cite{EDVACreport1945}.
	His publication is commonly referred to as "von Neumann architecture", although von Neumann in the first sentence makes clear that "\textit{The considerations which follow deal with the \textbf{structure} of a very high speed automatic digital computing system, and in particular with its \textbf{logical control}.}" He did \textit{not} define any architecture; just the opposite: he wanted to describe the operation of the engineering implementation in abstract terms, enabling mathematics to establish the science of computing. Moreover,
	"this is \dots the first substantial work \dots that clearly \textit{separated logic design from implementation}.  \dots 
	The computer defined in the 'First Draft' \textit{was never built, and its architecture and design seem now to be forgotten."}~\cite{GodfreyIEEE1993}.

	\subsection{The model of computing\label{sec:model}}
	
	"\textit{Devising a machine that can process information faster
		than humans has been a driving force in computing for
		decades}"~\cite{NeuromorphicSchumanSurvey:2017}.
	The model of computing did not change during the times, for centuries:
	\begin{itemize}
		\item the input operand(s) need to be delivered to the processing element
		\item the processing must be completely performed
		\item the output operand(s) must be delivered to their destination
	\end{itemize}
	The processing elements must be notified
	that all their operands are \textit{available}, and the result can be delivered to its destination
	only if the operation is \textit{completed}.
	Whether it is biological, neuromorphic, digital, or analog,
	the processing cannot even begin before its input operands are delivered to the place of operation, and vice versa; the output operand cannot be delivered until the computing process is terminated.
	\textit{Steps of computing logically depend on each other, and
		in their technical implementations they 
		need synchronization}.
	In this way, \textit{the data transfer and data processing block each other}.
	\textit{The total computing time of the model comprises both the transfer time(s) and the processing time. Even if, in the actual case, one of them can be neglected apart from the other.} 
	
	In most cases, simple operations are chained 
	(even if the same physical processing unit performs them), and several processing units must cooperate. That is, one's output is directed to
	another's input (say, logical gates are passing signals to each other
	or artificial/biological neurons send/receive "spikes"), but the above model must
	be kept in all cases. The chained processing units receive their input
	only later, when the unit they receive their input from, finished its operation.
	That is, \textit{any technological implementation converts the logical dependence of their operations to temporally consecutive phases}:
	\textit{the signal of one's output must reach the other's input before the chained computation can continue}.
	
	This limitation is valid for all implementations,
	from geared wheels to transporting electrons/ions.
	Electronic computers are no exception, although their operation is too fast to perceive with our human sensors. 
	This point deserves special attention when designing accelerators, feedbacks, and recurrent circuits: the computation considers the corresponding logical dependence through its timing relations.
	
	\subsection{The computer and the brain}
	
	Von Neumann discussed the operation of neurons in parallel with the
	operation of his intended technological implementation.
	In this context, it is clear that his model was \textit{biology-inspired},
	but (as can be concluded from the next section) because of its timing relations, it was not \textit{biology-mimicking}.
	It is just a joke of the technological development that our electronic computing devices today work in a more similar regime
	to that of our brain than vacuum tubes (and the mathematical abstraction based on it). A significant difference is that our brain works asynchronously, while our computing systems work (mostly) in a synchronized way.
	
	\subsection{Timing relations}

	In his famous publication~\cite{EDVACreport1945}, von Neumann made a careful feasibility analysis and warned: "\textit{6.3 At this point, the following observation is necessary. In the human nervous system, the conduction times [transmission times] along the lines (axons) can be longer than the synaptic delays [processing times], hence our above procedure of
		neglecting them aside from $\tau$ [the processing time] would be \textbf{unsound}.
		In the actually intended vacuum tube interpretation, however, this procedure is justified: $\tau$ is to be about a microsecond, an electromagnetic impulse travels in this time 300 meters, and as the lines are likely to be short compared to this, the conduction times may indeed be neglected. (It would take an ultra-high frequency device -- $\approx 10^{-8}$ seconds or less -- to vitiate this argument).}"

	Von Neumann was aware of the facts and the technical development: 
	"\emph{we will base our considerations on a hypothetical element, which functions essentially like a vacuum tube. \dots\ We reemphasize: This
		situation is only temporary, only a transient standpoint \dots After the
		conclusions of the preliminary discussion the elements will have to be
		reconsidered}"~\cite{vonNeumannOrigins}. 
	Unfortunately, the computing paradigm and the computer as a device were too successful, so the elements have never been reconsidered, and they were never updated according to the more advanced technology that appeared after his early passing.

	\subsection{The synchronous operating mode}
	
	As von Neumann explicitly warned in his "First Draft"~\cite{EDVACreport1945}, section 5.4, the operations must be synchronized appropriately, for example: "\textit{In division the calculation of a digit cannot even begin unless all digits to its left are already known}". That is, computing faces further limitations inside its technical implementation. The computation operation phases must be appropriately synchronized; furthermore, the parallelization must be carried out with care. We can add: as well as the acceleration of computations, including computing feedback and recurrent relations.
	
	The synchronization can be achieved by different means.
	The availability of the operand must be signaled, anyhow: either on a per-operand basis (asynchronous operation) or using some central control unit
	(synchronous operation). 
	The biology uses asynchronous mode:
	the spikes use self-timing (they carry also timing information).
	Just notice here that the technical delivery of 'spikes', due to the lack of
	real parallel connections suffer from distortions. 
	The technological imitation of biological synchronicity requires special care
	(time-stamping is not sufficient): an \textit{average} real-time simulation of the operation of a system does not necessarily mean the correct simulation of the \textit{individual biological events}.

	\subsection{The dispersion of the synchronization}
	
	In the same section 5.4, von Neumann told "\textit{We propose to use the delays $\tau$ as absolute units of time which can be relied upon to synchronize the functions of various parts of the device. The advantages of such an arrangement are immediately plausible}".
	When choosing such an absolute time unit, a "worst-case" timing must be chosen, that \textit{inherently introduces performance loss} for the "less wrong" cases.
	The technical systems, following von Neumanns proposal \textit{for the vacuum tube technology only}, typically use a central clock, and it is the designer's responsibility
	to choose a reasonable (but arbitrary) cycle period for the central clock.

	Using the model from section~\ref{sec:model}, we see that in chained computing operations \textit{the total execution
		times rather than the processing times} must be as uniform as possible: if either their 
	\textit{processing times} (different complexity) or \textit{transfer times} (different connection technology or just \textit{different signal propagation length}) differ,
	the synchronization inherently introduces some performance loss.
	The processing elements will be idle until the next clock pulse arrives
	if either their processing time is shorter or their data delivery time is longer. This difference in the arrival times is why von Neumann emphasized: "\textit{The emphasis is on the exclusion of a dispersion}"~\cite{EDVACreport1945}.
	Yes, his statement at the beginning of the section is true for the well-defined \textit{dispersionless synaptic delay $\tau$} he assumed, but not at all for today's processors. The recent activity to consider asynchronous operating modes~\cite{AsynchronParadigm:2013,FurberNeuralEngineering:2007,NeuromorphicSchumanSurvey:2017,PhysicsForNeuromorhicComputing:2020,BrainInspiredBlocks:2020}
	is motivated by admitting that \textit{the present synchronized operating mode is disadvantageous in the non-dispersionless world}.
	
	\section{Scrutinizing the dispersion}
	
	In von Neumann's abstraction a "\textit{well-defined dispersionless synaptic delay $\tau$ [processing time]}" is used.
	However, he used the word "dispersion" only in a broad (and only mathematical) sense, but he did not analyze its dependence on the actual physical conditions. Given that "\textit{The emphasis is on the exclusion of a dispersion}", 
	we define a merit for the dispersion, using the technical data of the given implementation. We provide a "best-case" and "worst-case" estimated value for the transfer time and define the dispersion as the
	geometric mean of the minimum and maximum "Proc transfer" times, divided by the processing time.
	
	\subsection{The case of EDVAC}
	
	Von Neumann mentioned that a "too fast" processor --with his words-- \textit{vitiates} his paradigm.
	If we consider a 300~$m^2$ sized computer room and the 3000 vacuum tubes estimated, von Neumann considered a distance between vacuum tubes about 30~cm as a critical value. At this distance, the transfer time is about three orders of magnitude lower than the processing time (neighboring vacuum tubes are assumed). The worst case is to transfer the signal to the other end of the computer room.
	With our definition, the dispersion of EDVAC is (at or below) 1~\%. 
	
	These limitations are why von Neumann justified his procedure \textit{for vacuum tube technology only}. He noted that
	using a hundred-fold higher frequency, even with vacuum tubes, vitiates the paradigm he proposed. 
	At such a frequency, the transfer time approaches the order of magnitude of the processing time, so neglecting the transfer time cannot anymore be justified: the \textit{apparent processing time}  (the clock time between consecutive computing operations) differs from the 
	\textit{physical processing time} by dozens of percents.

	\begin{figure}[t!]
		\includegraphics[width=1.\columnwidth]
		{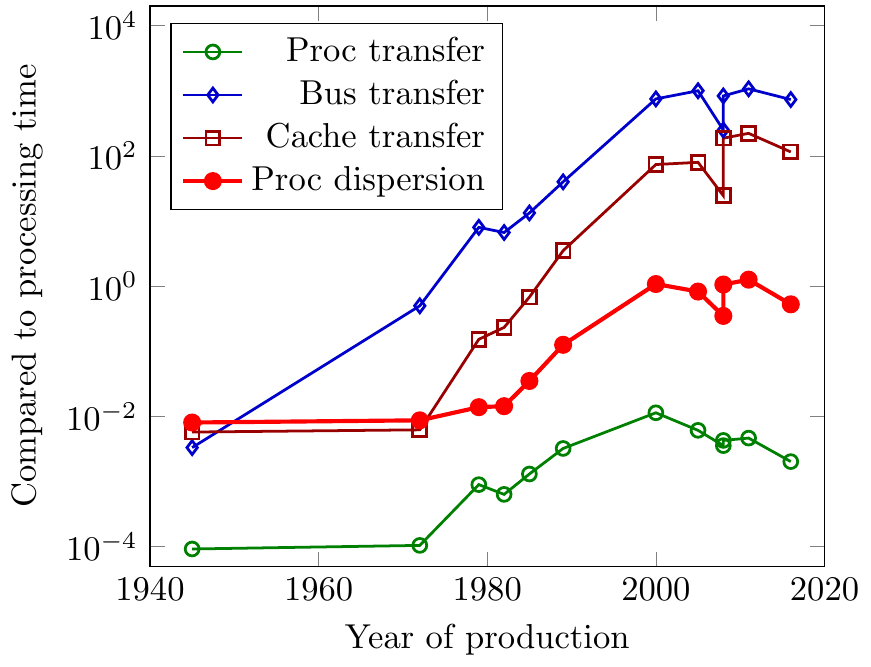}\vspace{-\baselineskip}
		\caption{The history of some relative temporal characteristics of processors, in function of their production year. Notice how cramming more transistors in a processor changed their temporal characteristics disadvantageously.\label{ProcessorDispersion}}
	\end{figure}

	\subsection{The case of integrated circuits}
	
	We derive dispersion for integrated circuits in the same way as discussed for vacuum tubes. Fig.~\ref{ProcessorDispersion} shows the dependence of different dispersion values on the year of fabrication of the processor. The technical data are taken from publicly available data\footnote{$https://en.wikipedia.org/wiki/Transistor\_count$} and from~\cite{EDVACEckertMauchly}.
	The figures of the merits are rough and somewhat arbitrary approximations. However, their consequent use enables us to draw limited validity conclusions without needing proprietary technological data.
	
	We estimate the distance between the processing elements in two different ways. We calculate the "average distance" of the transistors (the "best case") as the square root of the processor area divided by the number of transistors, and we consider it as a minimum distance the signals must travel between transistors\footnote{Notice that this transfer time also shows a drastic increase with the number of transistors, but alone does not vitiate the classic paradigm}. This value, divided by the clock frequency, is depicted as "Proc transfer" in Fig.~\ref{ProcessorDispersion}.   The maximum distance between the two farthest processing elements on the chip\footnote{Evidently, introducing clock domains and multi-core processors, shades the picture. However, we cannot provide a more accurate estimation without proprietary technological data}, which is the processor area's square root. 
	
	As the "Proc dispersion" diagram line shows, \textit{in today's technology the dispersion is near to unity}.
	That is, \textit{we cannot apply the "dispersionless" classic paradigm any more}\footnote{Reaching the plateau of the diagram lines coincides with introducing the "explicitly parallel instruction set computer"~\cite{EPIC:2000}:
		that was the maximum that the classic paradigm enabled.}. Furthermore, this means that the operating regime of our today's processors are more close to the operating regime
	of our brain (where explicit "spatiotemporal" behavior is considered and is "unsound" to use the classic paradigm to describe it) than the operating principle abstracted in the classic paradigm. 
	
	As we experience it, since the dispersion approached unity about two decades ago, only a small fragment of the input power can be used for computation; the rest of it is dissipated (produces only heat). \textit{The dispersion of synchronizing the computing operations} vastly increases the cycle time,  decreases the utilization of all computing units, and enormously increases the power consumption of computing~\cite{ClockDistribution:2012,DarkSilicon2012}. It is one of the major reasons for the inefficiency~\cite{InefficiencyHameed2010} of the ASIC circuits, and led to the symptom that moving data requires more energy~\cite{WhyNotExascale:2014} than manipulating it.
	The increased \textit{dispersion} enormously \textit{decreases performance} as the complexity and the \textit{relative transfer-to-processing time} increases.      
	
	\subsection{The case of technology blocks}
	
	Given that the processing elements and the storage elements usually are fabricated as separated technological blocks\footnote{This technological solution is misinterpreted as "the von Neumann architecture"; the \textit{structure} and \textit{architecture} are mismatched}, and they are connected by wires (aka bus), we also estimated a "bus transfer" time.
	The memory access time in this way is extended by the bus transfer time. We assumed that a cache memory is positioned at a(n average) distance of half processor size because of this effect. This time is shown as "Cache transfer" time. The cache memories appeared about the end of the 1980s, when it became evident that the bus transfer drastically increases the memory transfer time (cache data can be calculated for all processors, however).
	
	An interesting parallel is that both EDVAC and Intel 8008 have the same number of processing elements, and also the relative derived processor and cache transfer times are in the same order of magnitude.
	However, notice that the bus transfer time's importance has grown and started to dominate the single-processor performance in personal computers.
	A decade later, the physical size of the bus necessitated to introduce cache memories. 
	The physical size led to saturation in all relative transfer times.\footnote{The real cause of the "end of the Moore age" is that Moore's observation is not valid for the bus's physical size} The slight decrease in the relative times in the past years can probably be attributed
	to the sensitivity of our calculation method to the spread of multi-cores; this suggests to repeat our analysis method with proprietary technological data.

	\subsection{The need for communication}
	Also, at the time when von Neumann proposed his paradigm, there was literally only one processor.
	Because of this, questions how it will communicate and cooperate with the rest of (computing)
	world, should not be asked: there was no external world and no other processor.
	Today, billions of processors are fabricated in every year. They use ad-hoc methods and
	ideas about \textit{cooperation and communication}; furthermore, they use their \textit{payload time} for that activity.
	Furthermore, that activity increases the non-parallelizable portion of the tasks.
	It is very desirable to extend computing
	paradigm with considering the presence of other processors. 
	
	\subsection{Using new physical effect/ technology/ material in the computing chain}
	
	Given that the total execution time also comprises the transfer time,
	the meaningful analysis must consider the full time budget
	of the computing operations.

	\subsubsection{Using memristors for processing}
	A recently proposed idea is to replace slow digital processing with quick analog processing~\cite{RecipeMemristor:2020,NatureBuildingBrain:2020}, and \textit{may be proposed using any future new physical effect and/or material}. 
	Similar analysis can be followed in connection with the new physical effects considered in~\cite{RebootingComputingModels:2019}: they decrease the physical processing time, \textit{only}. To make them
	useful for computing, their in-component transmission time and especially the inter-component transmission time must be considerably decreased.
	
	It sounds good that "\textit{The analog memristor array is effectively the neural
		network laid out in the form of a crossbar, which can perform the
		entire operation \textbf{in one clock cycle}}"~\cite{BrainInspiredBlocks:2020}. In brackets, however fairly added, that \textbf{\textit{(not counting the clock cycles that
			may be required to fetch and store the input and output data)}}.
	Yes, the full story is that all operands of the memristor array 
	must be transferred to its input (and previously, must be produced),
	and the results must be transferred to their destination; usually
	through a (high speed) serial bus. The total time of the memristor-related operations shall be compared to the total time of conventional operations
	to make a fair comparison.
	
	\subsubsection{Half-length operands vs. double-length ones}
	The mutual locking of the transfer (and other, non-immediately payload operations) and the payload operations similarly leads to 
	disappointing efficiency improvement when one attempts to use
	half-length operands instead of double-length ones.
	The expectation behind the idea is that the shorter operand length
	may increase by a factor of four the desperately low efficiency
	of the artificial intelligence class applications running on supercomputers.
	One expects a four-fold performance increase when using half-precision rather than double-precision operands~\cite{HalfPrecisionArithmetic:2017},
	and the power consumption data underpin that expectation. However, the measured increase in computing performance was only three times higher: their temporal behavior limits the utility of using shorter operands, too.
	The housekeeping (such as fetching, addressing, incrementing, branching)
	remained the same, and because of the mutually blocking nature of the payload-non-payload operations, the increase of the payload performance
	is significantly lower. In the case of AI-type workload, the transfer time is significantly higher (the housekeeping can be neglected apart from transfer time), so the performance with half-precision and double precision operands differ only marginally. 
	For details, see~\textbf{\cite{VeghHowMany:2020,VeghScalingANN:2020}}.

	\subsubsection{The role of transfer time}
	
	The relative weight of the data transfer time has grown tremendously for many reasons.  Firstly, miniaturizing processors to sub-micron size while keeping the rest of the components (such as buses) above the centimeter scale. Secondly, single-processor performance has stalled~\cite{GameOverYelick:2011}, mainly because of reaching the limits,
	the laws of nature enable~\cite{LimitsOfLimits2014} (but, as we present, also because of tremendously extending its inherent idle waiting times). Thirdly, making truly parallel computers failed~\cite{AsanovicParallelCACM:2009}, and we can reach the needed high computing performance only through putting together an excessive number of segregated processors. This latter way replaced \textit{parallel computing} with \textit{parallelized sequential computing} (aka distributed computing),
	disregarding that the operating rules of the latter~\cite{ScalingParallel:1993}\textbf{\cite{VeghReevaluate:2020}\cite{VeghHowMany:2020}} sharply differ from those experienced with segregated processors.
	Fourthly, the utilization mode (mainly multitasking) forced us to use an OS, which imitates a "new processor" for a new task, at serious time expenses~\cite{Tsafrir:2007}\cite{armContextSwitching:2007}.
	Finally, the idea of "real-time connected everything"  introduced geographically large distances with their corresponding several millisecond data transfer times.
	Despite all of this, the idea of non-temporal behavior was confirmed by accepting the idea of "weak scaling"~\cite{Gustafson:1988}, 
	suggesting that \textit{all housekeeping times, such as organizing the joint work of parallelized serial processors, sharing resources, using exceptions and OS services, delivering data between processing units and data storage units, are negligible}.

	\subsubsection{How the presence of transfer time was covered}
	
	The experience showed that even within the core, wiring (and its related transfer time) has an increasing weight~\cite{LimitsOfLimits2014} in the timing budget.  
	When reaching the technology limit about $200~nm$
	around the year 2000\footnote{https://en.wikipedia.org/wiki/Transistor\_count}, wiring started to dominate 
	(compare this date to the year when saturation is reached in Fig.~\ref{ProcessorDispersion}). Further miniaturization can enhance the computing performance only marginally
	but increases the issues due to approaching the limiting interaction speed, as discussed below.

	It is accepted that logical gates have an
	operating time, and to compensate for the different number of gates
	on different paths, functionally not needed (such as invert-reinvert)
	gates are inserted on the path with fewer gates. The design comprises several clock domains, and some bigger parts of the design run with different synchronization.
	These solutions can more or less compensate for the evident time difference.
	However, on one side,
	computing systems "\textit{have a clock signal which is distributed in a tree-like fashion into every tiny part of the chip\dots Approximately 30~\% of the total consumption of a modern microprocessor is solely used for the clock signal distribution}."~\cite{ClockDistribution:2012} 
	It was the case 10 years ago, so for today, we can assume 50\% (and about the same amount of power is needed for cooling).
	On the other side, the difference in the length of physical signal paths
	causes a "skew" of the signals, which became a major challenge
	in designing high-performance systems. It leads to additional performance loss (mainly due to \textit{the dispersion of the clock signal needed to synchronize the computing operations}~\textbf{\cite{VeghMissingSecondDraft:2020}}).
	Even inside the die: the segregated processors have very low efficiency~\cite{InefficiencyHameed2010}. Despite this, today, wafer (and even multi-wafer~\cite{VerificationBrainScaleS:2020}) sized systems are also under design and in use.
	
	In complete systems, such as supercomputers running 
	HPCG workload, only 0.3~\% of the consumed energy goes for 
	computing, and this proportion is getting much worse if
	conventional systems attempt to mimic biology, such as to run an ANN
	workload. 
	The poor understanding of basic terms of computing resulted in that
	in supercomputing "\textit{the top 10 systems  [of TOP500] are unchanged from the previous list}"~\cite{Top500:2016}, and  that 
	"\textit{Core progress in AI has stalled in some quite different  fields}"~\cite{AIcoreProgressStalled:2020}; from brain simulation~\cite{NeuralNetworkPerformance:2018} to ANNs~\cite{DeepNeuralNetworkTraining:2016}; in general, the
	AI progress~\cite{AIcoreProgressStalled:2020} as a whole.

	\subsection{The case of non-negligible transfer time}

	At his time and in the age of vacuum tube technology, von Neumann did not feel the need to discuss what a procedure can justify describing the computing operation in a non-dispersionless case. However, \textit{he suggested reconsidering the validity of the neglections he used in his paradigm for any new future technology}. To have a firm computing paradigm for the present technologies, \textit{we need to consider the ratio of the transfer time to processing time}; we cannot neglect it anymore. The real question~\textbf{\cite{VeghMissingSecondDraft:2020} }is, the discussion of which is \textit{missing from the "First draft", what a procedure shall be followed if the transfer time is not negligible}?
	
	\section{The time-space system}
	
	\subsection{Considering the transfer time\label{sec:relativistic}}
	
	Although von Neumann explicitly mentioned that 
	the propagation speed of electromagnetic waves 
	limits the operating speed of the electronic components,
	until recently, that effect was not admitted in computing. 
	In contrast, in biology, the "spatio-temporal" behavior~\textbf{\cite{VeghBiologySpatioTemporal:2020}}
	was recognized very early.
	In both technical and biology-related computing,
	the recent trend is to describe computing systems theoretically and model their operation electronically using the computing paradigm proposed by von Neumann, which is undoubtedly not valid for today's technologies.
	Furthermore, as mentioned above, our computing devices' operating regime
	is closer to that of our brain than to the abstract model. That is, a similar description for both the computer and the brain would be adequate. 
	
	Fortunately, the spatio-temporal behavior suggests a "procedure" that can be followed in the case when the transfer time can even be longer than the processing time, the case which is missing from the "First Draft"~\textbf{\cite{VeghMissingSecondDraft:2020}}. Although biology --despite the name "spatio-temporal"--  describes the behavior of its systems
	using separated space and time functions (and as a consequence, needs ad-hoc suggestions and solutions for different problems), it has one common attribute with
	technical computing: in both of them, the information transfer speed is limited. 
	Because of its physical (and maybe philosophical) relevance, a more detailed explanation is given in~\textbf{\cite{VeghTemporal:2020}}.
	
	For the first look, it seems to be strange to describe such systems with (the inverse of) the Minkowski transform, given that it became famous
	in connection with Einstein's theory of special relativity. However, in its original form, only the existence of a limiting speed is assumed.
	This latter feature enables us to describe
	the correct behavior of information processing in science-based technological implementations and biology
	for any combination (ratio) of the transfer time and the processing time. The key idea is to transform the spatial distances between computing components (that can be $Si$ gates, cores, network nodes, biological or artificial neurons) to time (measured with the limiting speed along the signal path).

	We only assume that a limiting speed exists and that \textit{transferring information in the system needs time}. 
	In our approach, \textit{Minkowski provided a mathematical method to describe information transfer phenomena in a world, where the interaction speed is limited}.
	The only new assumptions we make are that the events also have a processing time, such as an atomic transition, executing a machine instruction, or issuing/receiving a neural spike; furthermore, the interaction speed is other than the speed of light.

	Special relativity describes the space around us with  \textit{four-dimensional space-time} coordinates, and \textit{calculates the fourth spatial coordinate from the time as the distance the light traverses in a given time}; simply because around us
	the distance was the easily accessible, measurable quantity -- hundred years ago.
	
	In computing, distances get defined during the fabrication of components and assembling the system.
	They may be different in different systems; however, they must meet their \textit{timing constraints}.
	In biological systems, nature defines neuronal locations and distances, and in 'wet' neuro-biology,
	\textit{signal timing rather than axon length is the right (measurable) parameter.} To introduce \textit{temporal logic} (meaning: the logical value of an expression depends on WHERE and WHEN is it evaluated) into computing, the reverse of the Minkowski transformation is required: the time, rather than the position, is the primary measurable quantity. 
	We need to use a special four-vector, where \textit{all coordinates are time values}:
	the first three are the corresponding local coordinates 
	(divided by the speed of interaction), having a time dimension.
	The fourth coordinate is the time itself.
	\textit{Distances from an event's location are measured along their access path; they are not calculated from their corresponding spatial coordinates}.
	
	In the case of electronics, 	the limiting speed is connected to the speed of light
	(in biological systems, the case is more complicated, but the speed of interaction is finite, anyhow).
	Given that we use the \textit{time} as the primary variable, the formalism can also be used to describe neuronal operation (where the conduction speed is modulated) with time, rather than space.  However, in the latter case, the formalism is less straightforward. At the same time, the famous hypothetical experiment can be excellently described in our \textit{time-space} system, too. With that experiment we demonstrate, that the solid mathematical background provided by the Minkowski-transform (and all associated behavior of modern science, also describing modern technological materials) is preserved. In the following chapters 
	we use the transformed coordinates only to describe the temporal behavior of computing systems.
	
	\textit{That is, the only change introduced to logic functions of computing is that
		they are not any more evaluated implicitly at point (0,0,0,0).
		Instead, they are evaluated at a point (x,y,z,t) of the time-space}.
	Below, we introduce the idea and its notations. Through describing the famous hypothetic experiment,
	we demonstrate that our \textit{time-space} system is equivalent to the commonly used \textit{space-time} systems. The validity and the mathematical features of the latter have been scrutinized exclusively in the past 140 years.
	So, on one side, we use the solid background the computer science is based upon: mathematics; on the other side, we extend it with the similarly solid background of \textit{time-space}
	formalism.

	\begin{figure}[t!]
		\includegraphics[width=1.\columnwidth]
		{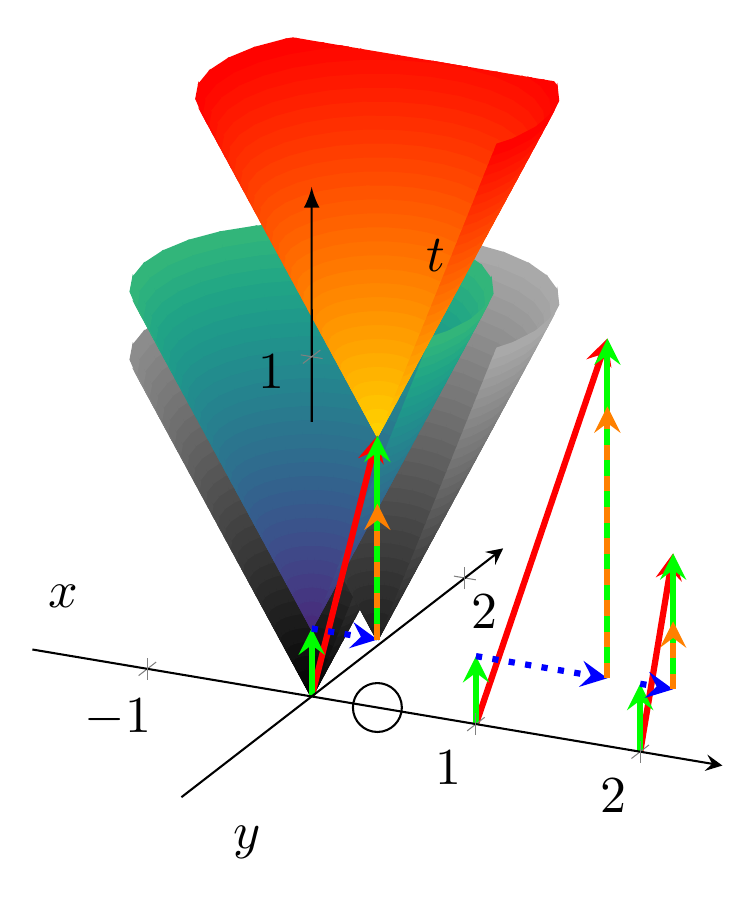}\vspace{-\baselineskip}
		\caption{The computing operation in \textit{time-space} approach. The processing operators can be gates, processors, neurons, or networked computers.\label{fig:RelativisticComputation}
			The "idle waiting time", rooting in the finite interaction speed and the physical distance of computing elements (see mixed-color vectors in figure), is one of the major sources of computing systems' inefficiency.}
	\end{figure}
	
	\subsection{Introducing the time-space system}
	
	With the reasoning above,   we introduce \textbf{\cite{VeghTemporal:2020}} a \textit{four dimensional time-space} system. The resemblance with the Minkowski space is evident,  and the name difference signals the different utilization methods.
	
	In our particular coordinate system (used in some figures below), formally (x,y,t) coordinates are used (for better visibility, the third spatial coordinate is omitted). What happens at time $t$ in a computing component at position (x,y) happens along a line parallel with axis $t$,  along a line perpendicular to plane at (x,y). 
	\textit{The objects are annotated with their spatial position coordinates 'x' and 'y',  but they are time values: how much time the signal having the limiting speed needs to reach that point.} 
	Several 'snapshots' taken at time $t$  about the same computing object are shown in the 3-D figures, on top of each other.
	The computing objects may alternatively be positioned at some arbitrary position that corresponds to the same \textit{time distance} from point (0,0,0) (a cylindrical coordinate system would be adequate but would make both visualization and calculations much harder to follow).
	
	The arrows in the same horizontal plane represent the same time (no transmission). 
	\textit{The interaction vectors are neither parallel with the time axis nor are in a spatial plane: both their temporal and spatial coordinates change as the interaction propagates. In electronic systems, the speed of interaction is constant,
		so the vectors are simple lines. Those spatial vectors are displayed with their different projections in the corresponding figures, enabling their easy trigonometric calculation. Their physical (spatial) length corresponds to the physical time between their endpoints}.
	The horizontal blue arrows are just helper lines: the position (annotated by x,y, but denoting the time the signal from (0,0,0) needs to reach this position) 
	is projected to \textit{time} axes $x$ and $y$ (the XY plane).
	The red arrow is the vectorial sum of the two projections, also in that plane.
	
	At the positions in the (x,y) plane, the events happen at the same time (although the other processing units will be
	able to notice that at a correspondingly later time, i.e., in another XY plane), and the events happening in connection with a processing unit, are aligned on an arrow parallel with axis~$t$. In this sense, \textit{we can interpret 'classic computing':
		our objects are compressed to one single XY plane: all events, at all objects, happen simultaneously.
		Even as the time distance between our computing objects is zero, because of the instantaneous transmission (independently from its technical implementation),
		the figure shall be represented with a mathematical point at (0,0,0).}
	The processing time is only an engineering imperfectness, according to the 'classic computing science'.
	
	\subsection{Validating the time-space system}
	Figure~\ref{fig:RelativisticComputation} (essentially a light cone in 2D space plus a time dimension) on one side demonstrates that our reversed Minkowski transform is equivalent for discussing the famous hypothetic experiment of relativity with the original transform, on the other side shows \textit{why time must be considered explicitly in all kinds of computing}. The figure shows that an event 
	appears in our \textit{time-space system} at point (0,0,0). The only (very plausible) difference to the classic hypothetic experiment is that we assume that signaling needs some time, too
	(not only the interaction, but also its technical implementation needs time).
	Our observers (fellow computing objects) are located on the 'x' axis; the vertical
	scale corresponds to time.
	In the classic physical hypothetical experiment, a light is switched on in the origo: the need to perform a calculation appears. The observers
	switch their light on (start their calculation) when they notice that
	the first light is switched on (the instruction/operand reaches their position). 
	The distance traveled by the light (the signal carrying the instruction) is given as the value of time multiplied by the speed of light (signal speed).
	At any point in time on the vertical axis, a circle describes the propagation of light (signal). In our (pseudo) 3-dimensional system, the temporal behavior is described as a conical surface, known as the \textit{future light cone} in science.
	
	Both light sources (computing objects) have some 'processing time',
	that passes between noticing the light (receiving an instruction/operand)  and switching their light (performing an instruction).
	An instruction is received at the bottom of the green arrow.
	The light goes on at the head of the arrow (i.e., at the same location, but at a later time)
	when the 'processing time' $T_p$ passed. Following that, the light propagates
	in the two spatial dimensions as a circle around the axis 't'.
	Observers at a larger distance notice the light at a later time:
	a 'transmission time' $T_t$ is needed.
	If the 'processing time' of the light source of our first event were zero,
	the light would propagate along the gray surface at the point (0,0,0).
	However, because of the finite processing time of the source, the signal  propagates along the
	blueish cone surface, at the head of the green arrow. 
	
	A circle marks the position of our other computing unit on the axis 'x'.
	With zero 'transmission time', a second gray conical surface (at the head of the horizontal blue dotted arrow)
	would describe the propagation of its signal. However, this second 'processing time' can only begin when our second processing unit receives the instruction at its position:
	when the mixed-color vertical dashed arrow hits the blueish surface.
	At that point begins the 'processing time' of our second processing unit;
	the yellowish conical surface, beginning at the second vertical green arrow, describes the second signal propagation.
	The horizontal (blue dotted) arrow describes the physical distance of the second computing unit (as a time coordinate) from the first one,
	the vertical (mixed color dashed) arrow describes the time delay of the instruction.
	It comprises two components: $T_t$ transmission time (mixed color)   to the observer  unit and 
	its $T_p$ processing time (green). The light cone of the observer (emitting the result of calculation) starts at $t=2*T_p+T_t$.
	
	The red arrow represents the resulting \textit{apparent processing time} $T_A$: 
	the longer is the red vector; the slower is the system.
	As the vectors are in the same plane,
	$T_A = \sqrt{T_t^2+(2\cdot T_p+T_t)^2}$, that is  $T_A = T_p\cdot \sqrt{R^2+(2+ R)^2}$.
	\textit{The apparent time is a non-linear function 
		of both of its component times} and \textit{their ratio $R$}.
	If more computing elements are involved, $T_t$ 
	denotes the longest transmission time. (Similar statement is valid if $T_p$ times are different.) The effect is significant:  if $R=1$, the apparent execution time of performing the two computations is more than three times longer than $T_p$.
	
	Two more observers are located on the axis 'x', at the same position, to illustrate the influence of transmission speed (and/or ratio $R$). For visibility, their timings are displayed at points '1' and '2', respectively. 
	In their case \textit{the transmission speed differs by a factor of two} compared to that displayed at point '0'; in this way
	three different $R=T_t / T_p$ ratios are used.
	Notice that at half transmission speed (the horizontal blue arrow is twice as long as the one at the origo)
	the vector is considerably longer, while at
	double transmission speed, the decrease of the time
	is much less expressed\footnote{This wants only to illustrate the effect of transmission speed on observations. This phenomenon is discussed in detail in~\textbf{\cite{VeghHowMany:2020}}.}. 
	
	\subsection{Scrutinizing the temporal behavior}
	
	The temporal behavior means that the value of logical functions depends on both time and place of 
	evaluating the corresponding logical function.

	Notice that some consequences are stemming immediately from the nature of our model. Now two computing elements are sitting 
	at point(0,0,0) and (0,1,0). The second element calculates something that expects the calculation of the first element as its input operand of the second calculation.
	Consider that the result is inside the green arrow during its processing time
	and comes to light after that time.
	As visible from the discussion and the figure, the first event happens at a well-determined position and time values in our \textit{space-time} system. Its spatial coordinates agree with those of the spatial position of the first element.
	Furthermore, they are different from those of the second element: the result must be transported:
	it shall be delivered to the position of the second processing unit.
	
	The event starts to propagate, and its final destination's coordinates differ both in time and space from those of the origin. It would be described as positioned at the head of the red vector
	(such vectors are neither vertical vectors, parallel with the $t$ axis,  nor lie within the plane (x,y) ).
	During this transfer, the space coordinate
	(the projection of the \textit{time-space} distance to the (x,y) plane)  changes to the coordinates of the head of the blue arrow: here \textit{was} the
	observer when the event happened.
	In the meantime, however, the time passed for the observer, and now the vertical arrow 
	at its position describes its coordinate.
	That vertical arrow comprises two contributions: the upper green arrow represents the processing time 
	of the observer, and we also have the length of the mixed-color arrow (the idle waiting): it has the same length as the blue dotted arrow:
	our observer must wait for such a long time to receive the signal. Given that all signals are in the plane (x,t), 
	the actual time distances can be calculated straightforwardly. 
	The projection of the event to axis $t$ is $T_p+T_i+T_p$,
	that is \textit{the apparent processing time includes idle time $T_i$, that equals 
		to the transfer time between the two processing units}.
	
	From our observation point, we see that the delivery times are considerably higher than those based on summing projections to the $t$ axis.
	Using different words, the apparent execution time in our a \textit{time-space system} is indeed more prolonged
	than the time difference calculated based on only the values projected to the $t$ axis. The difference (the length of the blue arrow) depends on the distance of the objects along the axis $x$ and the system's interaction speed. \textit{The figure suggests that the difference is the larger,
		the more complex operations are involved, and the larger is the physical size of the system}
	(provided that the system defines the interaction speed). 
	The primary result of that transformation is that \textit{the apparent processing time
		can be calculated by using simple geometrical relations}.
	
	When looking at the events from the direction of the $x$ axis in the (x,t) plane, we see that the
	total time corresponds to
	the sum of the two processing times in' classic computing'. The waiting time of the second unit is only
	an arrowhead (instant interaction). It denotes the corresponding logical dependence but does not increase processing time.
	The second unit must wait for the result of the first calculation because of their logical dependence.
	However, because of instant delivery, no additional waiting is required.
	Consequently, according to the "timeless (classical) paradigm, " observers' distance seems to be zero.
	The introduced extra dimension, \textit{time of interaction}, changes the picture drastically.
	The closer is the difference between the summed length of the two green vectors to the red vector's length; the larger is the speed of their interaction.
	\textit{They are equal, however, only if the interaction speed is infinitely large.}
	
	\subsection{Computing efficiency as a consequence of  temporal behavior}
	It has been discussed virtually infinite times in computing that the computing performance depends on many factors. However, the temporal dependence has never been included.
	As the discussion above suggested, time has a decisive role in computing;
	the \textit{idle waiting times} indeed degrade performance,
	so the rest of the discussion shall focus on the role of time in shaping performance.
	
	Given that the spatial distance is equivalent to time, they have a similar role.
	The processor operation itself has some idle time, and, as discussed in the next sections,
	the outdated technical principles and solutions add more (and slightly different) idle times.
	Moore's observation is valid only for the gate's density inside chips, but not for the components 
	(such as buses) connecting them, so the idle (transmission) time increases.
	\textit{This relation has a self-exciting effect: the low efficiency (that decreases as the required performance increases)
		means that a larger portion of energy consumption is used for heating (and because of that: needs more cooling),
		needing more physical space, which increases the physical distance of components, causing worse performance, and so on}.
	
	On one side, the communication between processors is implemented in a way that increases the non-payload, sequential-only
	portion of the task. On the other side, the physical time of transmission (that depends on both the speed of interconnection and the physical distance between the corresponding computing objects)
	also significantly contributes to degrading the efficiency.
	Also, the algorithms, how the components are used, and the parameters of the architecture all
	impact the system's computing efficiency. \textit{The benchmark data of the components define the hard limits, and their way of cooperation the soft limits that can be experienced.} That is, \textit{the temporal behavior of their components is a vital feature of computing systems, especially of the excessive ones, mainly if they target high computing performance,
		especially if they are running very demanding workloads}. 
	
	\section{Technical solutions for the vacuum-tube age}
	
	Some of the "classic" technical implementations,
	that -- due to the incremental development -- survived their technical 
	state-of-the-art, and (especially in the light of temporal analysis) need drastic revision.
	Also, research is going on (both in science and technology, with vast investments in the background) to find new materials/effects/technologies.
	However, science severely limits their usability: the temporal analysis provides a helping hand also here, enabling optimizing performance/cost.
	"Reinventing electronics"\cite{PhysicsForNeuromorhicComputing:2020} is necessary not only for computing devices but also their interconnection and modes of utilization.

	\begin{figure*}[t!]
		\includegraphics[width=1.\textwidth]
		{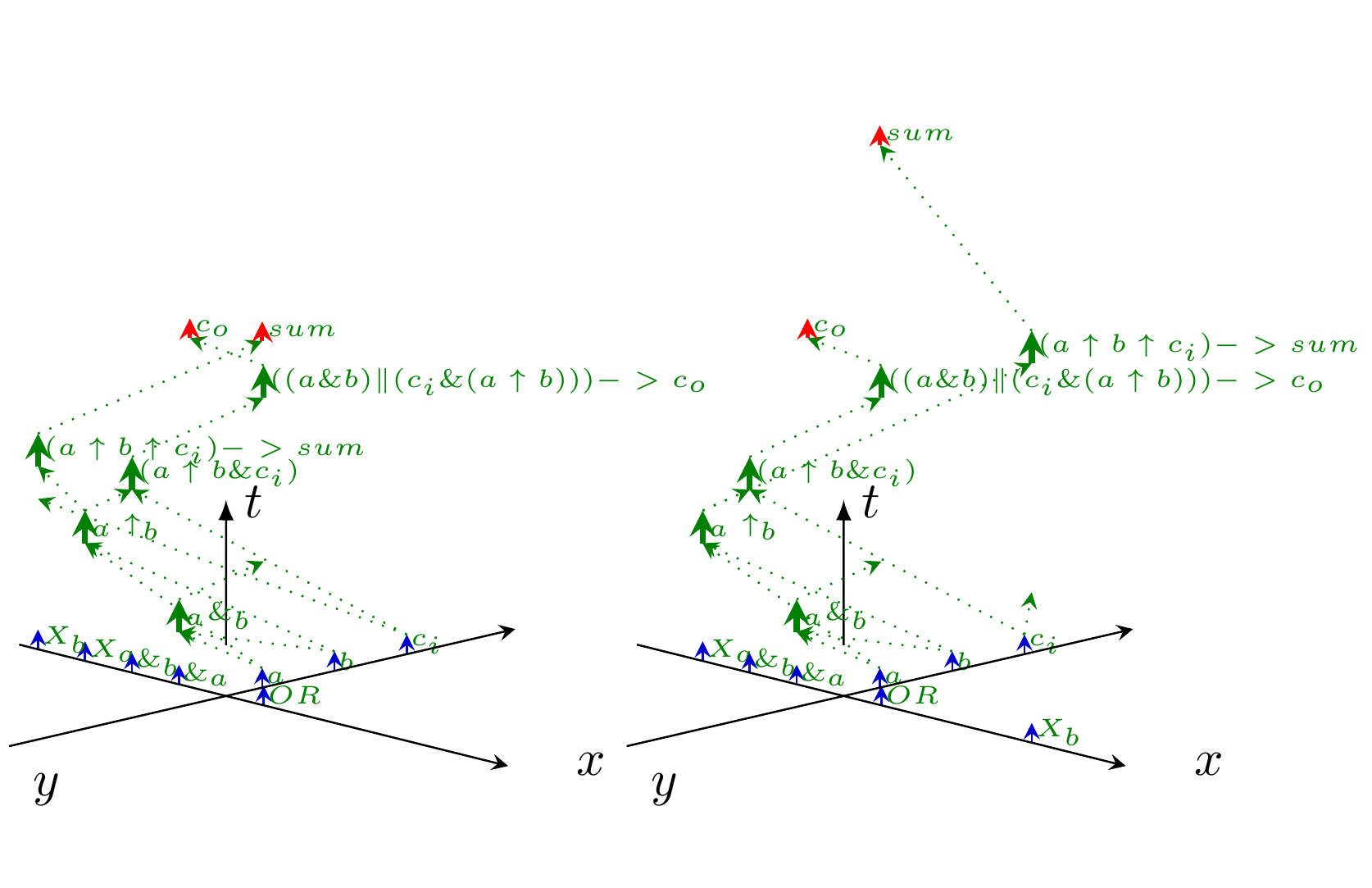}\vspace{-\baselineskip}
		
		\caption{The temporal dependence diagram of a 1-bit adder. 	The diagram shows the logical equivalent of the SystemC source code of Listing~\ref{lst:OneBitAdder},
			\textit{the lack of vertical arrows} signals "idle waiting" time (undefined gate output)
			Left side: the second XOR gate is at (-1,0).
			Right side: the second XOR gate is at (+1,0).
			Notice how changing the position of a gate affects signal timing.
			\label{fig:Adder1bitDouble}}
	\end{figure*}

	\subsection{Method of identifying bottlenecks of computing }

	\textit{The transmission time $T_t$  is an 'idle time'} from the point of view of computing:
	the component is ready to run, takes power, but does no useful work.
	Due to their finite physical size and the finite interaction speed (both neglected in the classic paradigm), \textit{the temporal operation of computing systems 
		results inherently in an idle time of the Processing Units, that \textit{can be a significant contributor to
			non-payload portion of their processing time}.
		Given that the different apparent processing times inevitably increase the dispersion, they can be a key factor of the experienced inefficiency of general-purpose chips~\cite{InefficiencyHameed2010}}. 
	With other major contributors, originating from the technical implementation of their cooperation,
	these "idle waiting" times sharply decrease the payload performance of computing systems. It is worth to discuss the inside-component and inter-component contributions 
	separately.
	
	In the spirit of the temporal behavior, we can set up two general classes of processing:
	the \textit{payload processing}  $T_p$ makes our operations directly related to our goal of computation;
	all other processing is counted as \textit{ non-payload processing}.  \textit{The merit we use is the time spent with that processing}. As will be shown, some portion of the non-payload processing time
	roots in laws of nature: \textit{computing inherently includes idle times}, some other portion (such as housekeeping) is not \textit{directly} useful. However, any kind of processing takes time and consumes energy.
	The task of designing our computing systems is to reduce the total processing time,
	i.e., \textit{to develop solutions that minimize the proportion of the 'idle' activity; 
		not only at a component level but also at the system level.} Scrutinizing the temporal diagrams of components, solutions, and principles is an excellent tool to find bottlenecks.
	
	In the figures below, near to (vertical) axis \textit{t} are shown vertical arrows (where payload processing happens) or lack of arrows (when non-payload processing occurs).  The large amount of non-payload processing 
	(that increases with system complexity) explains the experienced low computing efficiency of computing systems using those technical implementations.
	The proportions of times are chosen for better visibility and call attention to its effect rather than reflect some realistic arrangements.
	
	\subsection{Gate-level processing}\label{sec:Technical:Gates}
	
	Although for its end-users, the processor is the "atomic unit" of processing\footnote{The reconfigurable computing, with its customized processors and non-processor-like processing units, does not
		change the landscape significantly}, principles of computing are valid also 
	at "sub-atomic" level, at the level of gate operations.
	Describing the temporal operation at gate level is an excellent example, that \textit{the line-by-line compiling
		(sequential programming, called also Neumann-style programming~\cite{BackusNeumannProgrammingStyle}),
		formally introduces only logical dependence, but through its technical implementation
		it implicitly and inherently introduces a temporal behavior, too}.

	The one-bit adder is one of the simplest circuits used in computing.
	Its typical implementation comprises 5 logic gates, 3 input signals, and 2 output signals.  Gates are logically connected internally: they provide input and
	output for each other. 
	The relevant fraction of the equivalent source code is shown in Listing~\ref{lst:OneBitAdder}.
	
	\begin{lstlisting}[float,caption=The essential lines of source code of the one-bit adder implemented in 
	SystemC,label=lst:OneBitAdder]
	//We are making a 1-bit addition
	aANDb = a.read() & b.read();
	aXORb = a.read() ^ b.read();
	cinANDaXORb = cin.read() & aXORb;
	
	//Calculate sum and carry out
	sum = aXORb ^ cin.read();
	cout = aANDb | cinANDaXORb;
	\end{lstlisting}
	
	Fig.~\ref{fig:Adder1bitDouble}  shows the timing diagram of a one-bit adder, implemented 
	using common logic gates.  The three input signals are aligned on axis $y$; the five logic gates are aligned on axis $x$. Gates are ready to operate, and signals are ready to be processed (at the head of the blue arrows). The logic gates have the same operating time (the length of green vectors); their access time distance includes the needed multiplexing. The signals must reach their destination gate (dotted green arrows),
	that (after its operating time passes) produces the output signal, that starts immediately towards the next gate. The vertical green arrows denote gate processing
	(one can project the arrow to axis $x$ to find out the gate's ID),
	labeled with the produced signal's name.
	Because of the non-synchronized operating mode,
	there are "pointless" arrows in the figure. For example, signal $a\&b$
	reaches the $OR$ gate much earlier than the signal to its other input. Depending on the operands of $OR$,
	it may or may not result in the final sum. The gates always have an output signal, even if they did not receive their expected input.
	
	Notice that considering  the physical distance
	and the finite interaction speed drastically changes the picture we have (based on 
	"classic computing"), that the operating time of the adder is simply
	the sum of the "gate times".
	For example, the very first AND and XOR operations could work in parallel
	(at the same time), but the difference in their physical distance the signals must travel,
	changes the times when they can operate with their signals.
	Also, compare the temporal behavior of the signal $sum$
	on the two figures.
	The only difference between subfigures is
	that the second XOR gate moved to another place.

	\begin{figure}[t!]
		\includegraphics[width=1.\columnwidth]
		{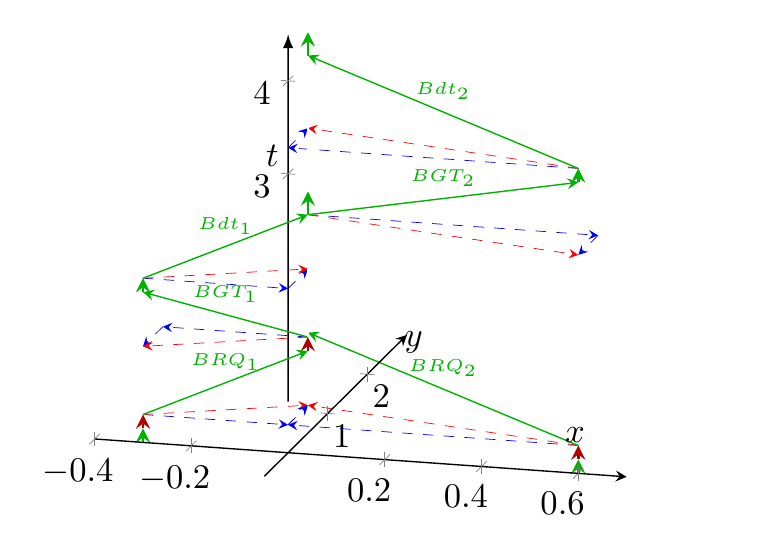}\vspace{-\baselineskip}
		\caption{The computing operation in \textit{time-space} approach. The processing operators can be gates, processors, neurons, or networked computers.\label{fig:Relativisticbus}
			The "idle waiting time", rooting in the finite interaction speed and the physical distance of computing elements (see mixed-color vectors in figure), is one of the major sources of computing systems' inefficiency.}
	\end{figure}

	\subsection{The serial bus}\label{sec:Technical:SerialBus}
	Fig.~\ref{fig:Relativisticbus} discusses, in terms of "temporal logic": why using high-speed buses for connecting
	modern computer components leads to very severe performance loss, especially when one
	attempts to imitate neuromorphic operation. The processors are positioned
	at (-0.3,0) and (0.6,0). The bus is at a position (0,0.5). The two processors make their computation 
	(green arrows at the position of processors), then they want to deliver their result to its destination. We assume that they want to communicate simultaneously.
	First, they must have access to the shared bus (red arrows). 
	The core at (-.3,0) is closer to the bus, so its request is granted.
	As soon as the grant signal reaches the requesting core, the bus operation is initiated, and the data starts to travel to the bus. As soon as it reaches the bus, the bus's high speed forwards it, and at that point, the bus request of the other core is granted, and finally, the calculated result of the second core is bused. 
	
	At this point comes into picture the role of the workload on the system: we presumed that the two cores want to use the single shared bus, at the same time, for communication. Given that they must share it, 
	\textit{the apparent processing time is several times higher than the physical processing time.
		Moreover, it increases linearly with the number of cores connected to the bus}, if a single high-speed bus is used.
	\textit{In vast systems, especially when attempting to mimic neuromorphic workload, 
		the bus's speed is getting marginal}.
	Notice that the times shown in the figure are not proportional: 
	the (temporal) distance between cores is in the several picoseconds range,
	while the bus (and the arbiter) are at a distance well above nanoseconds, so \textit{the actual temporal behavior (and the idle time stemming from it) is much worse than the figure suggests}.
	This behavior is why \textit{"The idea of using the popular shared bus to implement the communication medium is no longer acceptable, mainly due to its high contention."}~\cite{ReconfigurableAdaptive2016}.
	For a more detailed analysis see~\textbf{\cite{VeghTemporal:2020}}, and specifically for the case of artificial neural networks\textbf{~\cite{VeghScalingANN:2020}}.
	The figure suggests using another design principle instead of using the bus
	exclusively, directly from the computing component's position.
	
	\textit{When using a shared bus, increasing either the processing
		speed or the communication speed, alone, does not affect \textit{linearly}
		the total execution time any more. 
		Furthermore, it is not the bus speed that limits performance.}
	A relatively small increase in the transfer time $T_t$
	can lead to a relatively large change in the value of the apparent processing time $T_A$.
	This change leads to an incomprehensible slowdown of the system: its slowest component defines its efficiency. 
	The conventional method of communication may work fine as long as there is no competition
	for the bus, but leads to queuing of messages in the case of 
	(more than one!) independent sources of communication.
	The effect is topped by the bursty nature of communication caused by the need for central synchronization,
	leads to a "communicational collapse"~\cite{CommunicationCollapse:2018},
	that denies huge many-processor systems, especially neuromorphic ones~\cite{SpiNNaker:2013}.

	\begin{figure}[t!]
		\includegraphics[width=1.\columnwidth]
		{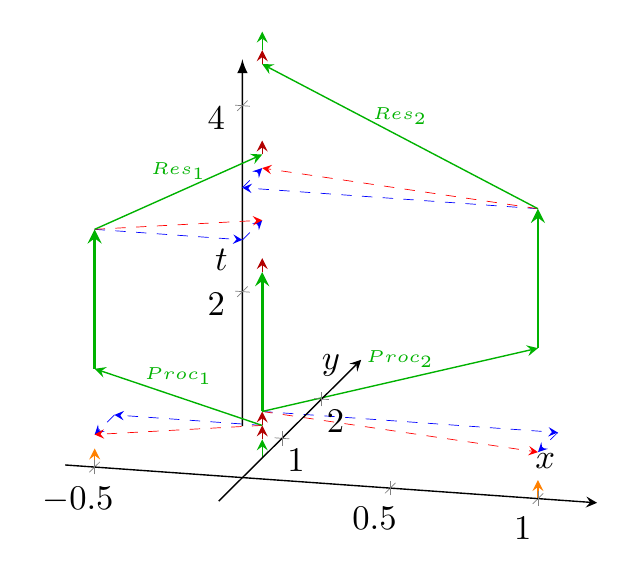}\vspace{-\baselineskip}	
		\caption{The operation of the "parallelized sequential" processing, in the time-space coordinate system. Parallel with axis \textit{t}, \textit{the lack of vertical arrows} signals "idle waiting" time, both for the coordinator and the fellow processors\label{fig:RelativisticDistributed}}
	\end{figure}
	
	\subsection{Distributed processing\label{sec:}}
	Given that single-processor performance stalled~\cite{ComputingPerformanceBook:2011}, 
	and building parallel computers failed~\cite{AsanovicParallelCACM:2009}, to reach the needed high
	computing performance, the computing tasks must be cut into pieces and be distributed between independently working single processors. Cutting and re-joining pieces, however,
	needs efforts both from programming and technology. The technology, optimized for solving
	single-thread tasks, hits back when several processors must cooperate, as cooperation
	and communication are not native features of segregated processors. The task was so hard that the famous Gordon Bell Prize~\cite{GordonBellPrize:2017} was awarded initially to achieve an at least 200-fold performance gain via distributing a task between several (even thousands of) processors.
	
	Fig~\ref{fig:RelativisticDistributed} depicts the time diagram of distributed parallel processing.
	One of the Processing Units (in our case, the one at (0,0.5)) orchestrates the operation,
	including receiving the start command and sending the result.
	This core makes some sequential operations (such as initializing data structures, short green arrow),
	then it sends the start command and operands to fellow cores at (-0.5,0) and (1,0), one after the other.
	Signal propagation takes time (depending on distance from the coordinator). After that time, fellow cores can start their calculation (their part of the parallelized portion).
	Of course, the orchestrator  Processing Unit must start all fellow Processing Units (red arrows), then it can start
	its portion of distributed processing. 

	\begin{figure}[t!]
		\includegraphics[width=1.\columnwidth]
		{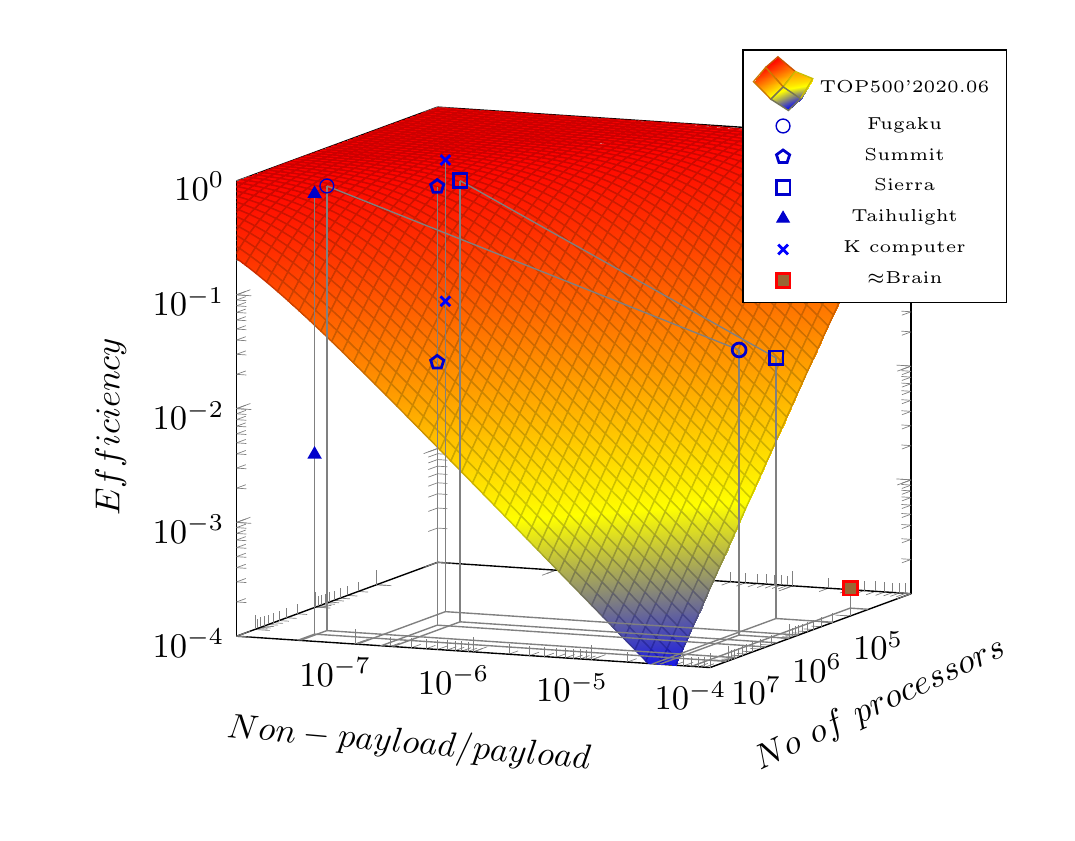}\vspace{-\baselineskip}	
		\caption{The 2-parameter efficiency surface (in function of the parallelization efficiency measured by benchmark HPL and the number of the PUs) as concluded from Amdahl's Law (see~\textbf{\cite{VeghHowMany:2020}}), in the first order approximation. Some sample efficiency values for
			some selected supercomputers are shown, measured with benchmarks HPL and HPCG, respectively.      Also, the estimated efficacy of brain simulation using conventional computing is shown.    \label{fig:EffDependence2020Log}}
	\end{figure}

	As the fellow Processing Units finish their portion, they must transmit their  data  $Res_i$  to the orchestrator,
	that receives those data in \textit{sequential} mode, and finally makes some closing \textit{sequential} processing.
	Again, the \textit{inherently sequential-only portion}~\cite{InherentSequentiality:2012} of the task increases with number of cores and its \textit{idle waiting time} (time delay of signals) increases with physical size (cable length).
	Notice also that the orchestrating Processing Unit must wait for the results from all fellow cores,
	i.e., \textit{the slowest branch defines performance}.
	This aspect is significant, if the physical transmission times of those units differ 
	(speed or distance is different), the task is not adequately split into equal portions.
	Modern hardware has undeterministic behavior~\cite{PerformanceCounter2013}\textbf{\cite{Vegh:2017:AlphaEff}},
	or the units are connected through an interconnection with stochastic behavior~\cite{PricePerformanceClouds:2017}.
	The times shown in the figure are not proportional and largely depend on the type of the system.
	For a detailed discussion, see~\textbf{\cite{VeghHowMany:2020}}.
	
	The dispersion in distributed systems is dramatically increased
	because of the physical distance of the computing components. The increase
	depends on the weight of the critically large distances.
	As analyzed in detail in~\textbf{\cite{VeghHowMany:2020}}, the different communication intensity increases the weight of the operation having
	a considerable temporal distance. For vast distributed systems, see Fig.\ref{fig:EffDependence2020Log},
	the efficiency sharply depends on the number of parallelized units and the goodness of their parallelization. "\textit{This decay in performance is not a fault of the
		architecture, but is dictated by the limited parallelism}"~\cite{ScalingParallel:1993}

	\section{Conclusion}
	The technological development made the neglections used in the commonly used computing model outdated and forces us to consider the original model, with correct timing relations. The stealthy nature of the development led to many technological implementations, including the synchronous operation and the high-speed bus that are not usable anymore. One obvious sign is the enormous dissipation of our computing systems, which is a direct consequence of the fact that the computing systems' dispersion is well above the theoretically acceptable level. 
	"Reinventing electronics"~\cite{PhysicsForNeuromorhicComputing:2020} is a must,
	and not only for building neuromorphic computing devices. The computing model is perfect, but the classic paradigm is used far outside its validity range. For today's technological conditions, the needed
	"contract"~\cite{AsanovicParallelCACM:2009} between mathematics and electronics, should be based on a paradigm that 
	considers the transfer time.

	\ifCLASSOPTIONcompsoc
	\section*{Acknowledgments}
	\else
	\section*{Acknowledgment}
	\fi

	Project no. 136496 has been implemented with the support provided from the National Research, Development and Innovation Fund of Hungary, financed under the K funding scheme. 
	
	\ifCLASSOPTIONcaptionsoff
	\newpage
	\fi


\begin{thebibliography}{10}
	\providecommand{\url}[1]{#1}
	\csname url@samestyle\endcsname
	\providecommand{\newblock}{\relax}
	\providecommand{\bibinfo}[2]{#2}
	\providecommand{\BIBentrySTDinterwordspacing}{\spaceskip=0pt\relax}
	\providecommand{\BIBentryALTinterwordstretchfactor}{4}
	\providecommand{\BIBentryALTinterwordspacing}{\spaceskip=\fontdimen2\font plus
		\BIBentryALTinterwordstretchfactor\fontdimen3\font minus
		\fontdimen4\font\relax}
	\providecommand{\BIBforeignlanguage}[2]{{%
			\expandafter\ifx\csname l@#1\endcsname\relax
			\typeout{** WARNING: IEEEtran.bst: No hyphenation pattern has been}%
			\typeout{** loaded for the language `#1'. Using the pattern for}%
			\typeout{** the default language instead.}%
			\else
			\language=\csname l@#1\endcsname
			\fi
			#2}}
	\providecommand{\BIBdecl}{\relax}
	\BIBdecl
	
	\bibitem{EDVACreport1945}
	J.~von Neumann, ``{First Draft of a Report on the EDVAC},''
	\url{https://web.archive.org/web/20130314123032/
		http://qss.stanford.edu/~godfrey/vonNeumann/vnedvac.pdf}, 1945.
	
	\bibitem{ComputingPerformanceBook:2011}
	S.~H. Fuller and L.~I. Millett, Eds., \emph{{The Future of Computing
			Performance: Game Over or Next Level?}}\hskip 1em plus 0.5em minus
	0.4em\relax National Academies Press, Washington, 2011.
	
	\bibitem{AsanovicParallelCACM:2009}
	K.~Asanovic, R.~Bodik, J.~Demmel, T.~Keaveny, K.~Keutzer, J.~Kubiatowicz,
	N.~Morgan, D.~Patterson, K.~Sen, J.~Wawrzynek, D.~Wessel, and K.~Yelick, ``{A
		View of the Parallel Computing Landscape},'' \emph{Comm. ACM}, vol.~52,
	no.~10, pp. 56--67, 2009.
	
	\bibitem{SoOS:2010}
	{S(o)OS~project}, ``Resource-independent execution support on exa-scale
	systems,'' \url{http://www.soos-project.eu/index.php/related-initiatives},
	2010.
	
	\bibitem{DeBenedictis_supercomputing:2014}
	\BIBentryALTinterwordspacing
	{Machine Intelligence Research Institute}, ``{Erik DeBenedictis on
		supercomputing},'' 2014. [Online]. Available:
	\url{https://intelligence.org/2014/04/03/erik-debenedictis/}
	\BIBentrySTDinterwordspacing
	
	\bibitem{TrueNorth:2016}
	J.~{Sawada~et~al}, ``{TrueNorth Ecosystem for Brain-Inspired Computing:
		Scalable Systems, Software, and Applications},'' in \emph{SC '16: Proceedings
		of the International Conference for High Performance Computing, Networking,
		Storage and Analysis}, 2016, pp. 130--141.
	
	\bibitem{GodfreyIEEE1993}
	M.~D. Godfrey and D.~F. Hendry, ``{The Computer as von Neumann Planned It},''
	\emph{IEEE Annals of the History of Computing}, vol.~15, no.~1, pp. 11--21,
	1993.
	
	\bibitem{NeuromorphicSchumanSurvey:2017}
	C.~D.~S. et~al, ``{A Survey of Neuromorphic Computing and Neural Networks in
		Hardware},'' 2017.
	
	\bibitem{vonNeumannOrigins}
	W.~Aspray, \emph{{John von Neumann and the Origins of Modern Computing}}.\hskip
	1em plus 0.5em minus 0.4em\relax MIT Press, Cambridge, 1990, pp. 34--48.
	
	\bibitem{AsynchronParadigm:2013}
	\BIBentryALTinterwordspacing
	S.~K. et~al, ``Acceleration of an asynchronous message driven programming
	paradigm on ibm blue gene/q,'' in \emph{2013 IEEE 27th International
		Symposium on Parallel and Distributed Processing}.\hskip 1em plus 0.5em minus
	0.4em\relax Boston: IEEE, 2013. [Online]. Available:
	\url{https://https://ieeexplore.ieee.org/abstract/document/ 6569854}
	\BIBentrySTDinterwordspacing
	
	\bibitem{FurberNeuralEngineering:2007}
	{Steve Furber and Steve Temple}, ``{Neural systems engineering},'' \emph{{J. R.
			Soc. Interface}}, vol.~4, pp. 193--206, 2007.
	
	\bibitem{PhysicsForNeuromorhicComputing:2020}
	D.~Markovic, A.~Mizrahi, D.~Querlioz, and J.~Grollier, ``{Physics for
		neuromorphic computing},'' \emph{Nature Reviews Physics}, vol.~2, pp.
	499--510, 2020.
	
	\bibitem{BrainInspiredBlocks:2020}
	J.~D. Kendall and S.~Kumar, ``{The building blocks of a brain-inspired
		computer},'' \emph{Appl. Phys. Rev.}, vol.~7, p. 011305, 2020.
	
	\bibitem{EDVACEckertMauchly}
	J.~J.~P.~Eckert and J.~W. Mauchly, ``{Automatic High-Speed Computing: A
		Progress Report on the EDVAC},'' Moore School Library, University of
	Pennsylvania, Philadephia, Tech. Rep. Report of Work under Contract No.
	W-670-ORD-4926, Supplement No 4, September 1945.
	
	\bibitem{EPIC:2000}
	M.~Schlansker and B.~Rau, ``{EPIC: Explicitly Parallel Instruction
		Computing},'' \emph{Computer}, vol.~33, no.~2, pp. 37--45, Feb 2000.
	
	\bibitem{ClockDistribution:2012}
	R.~Waser, Ed., \emph{{Advanced Electronics Materials and Novel Devices}}, ser.
	Nanoelectronics and Information Technology.\hskip 1em plus 0.5em minus
	0.4em\relax Wiley, 2012.
	
	\bibitem{DarkSilicon2012}
	H.~Esmaeilzadeh, E.~Blem, R.~St.~Amant, K.~Sankaralingam, and et~al., ``{Dark
		Silicon and the End of Multicore Scaling},'' \emph{IEEE Micro}, vol.~32,
	no.~3, pp. 122--134, 2012.
	
	\bibitem{InefficiencyHameed2010}
	\BIBentryALTinterwordspacing
	R.~Hameed, W.~Qadeer, M.~Wachs, O.~Azizi, A.~Solomatnikov, B.~C. Lee,
	S.~Richardson, C.~Kozyrakis, and M.~Horowitz, ``Understanding sources of
	inefficiency in general-purpose chips,'' in \emph{Proceedings of the 37th
		Annual International Symposium on Computer Architecture}, ser. ISCA
	'10.\hskip 1em plus 0.5em minus 0.4em\relax New York, NY, USA: ACM, 2010, pp.
	37--47. [Online]. Available: \url{http://doi.acm.org/10.1145/1815961.1815968}
	\BIBentrySTDinterwordspacing
	
	\bibitem{WhyNotExascale:2014}
	\BIBentryALTinterwordspacing
	H.~Simon, ``{Why we need Exascale and why we won't get there by 2020},'' in
	\emph{Exascale Radioastronomy Meeting}, ser. AASCTS2, 2014. [Online].
	Available: \url{https://www.researchgate.net/publication/261879110\\
		\_Why\_we\_need\_Exascale\_and\_why\_we\_won't\_get \_there\_by\_2020}
	\BIBentrySTDinterwordspacing
	
	\bibitem{RecipeMemristor:2020}
	\BIBentryALTinterwordspacing
	E.~Chicca and G.~Indiveri, ``{A recipe for creating ideal hybrid
		memristive-CMOS neuromorphic processing systems},'' \emph{Applied Physics
		Letters}, vol. 116, no.~12, p. 120501, 2020. [Online]. Available:
	\url{https://doi.org/10.1063/1.5142089}
	\BIBentrySTDinterwordspacing
	
	\bibitem{NatureBuildingBrain:2020}
	\BIBentryALTinterwordspacing
	``{Building brain-inspired computing},'' \emph{Nature Communications}, vol.~10,
	no.~12, p. 4838, 2019. [Online]. Available:
	\url{https://doi.org/10.1038/s41467-019-12521-x}
	\BIBentrySTDinterwordspacing
	
	\bibitem{RebootingComputingModels:2019}
	P.~C. et~al., ``{Rebooting Our Computing Models},'' in \emph{Proceedings of the
		2019 Design, Automation \& Test in Europe Conference \& Exhibition
		(DATE)}.\hskip 1em plus 0.5em minus 0.4em\relax IEEE Press, 2019, pp.
	1469--1476.
	
	\bibitem{HalfPrecisionArithmetic:2017}
	A.~Haidar, P.~Wu, S.~Tomov, and J.~Dongarra, ``{Investigating Half Precision
		Arithmetic to Accelerate Dense Linear System Solvers},'' in
	\emph{{Proceedings of the 8th Workshop on Latest Advances in Scalable
			Algorithms for Large-Scale Systems}}, ser. ScalA '17.\hskip 1em plus 0.5em
	minus 0.4em\relax New York, NY, USA: ACM, 2017, pp. 10:1--10:8.
	
	\bibitem{VeghHowMany:2020}
	\BIBentryALTinterwordspacing
	J.~V{\'{e}}gh, ``Finally, how many efficiencies the supercomputers have?''
	\emph{The Journal of Supercomputing}, vol.~76, no.~12, pp. 9430--9455, feb
	2020. [Online]. Available:
	\url{http://link.springer.com/article/10.1007/s11227-020-03210-4}
	\BIBentrySTDinterwordspacing
	
	\bibitem{VeghScalingANN:2020}
	\BIBentryALTinterwordspacing
	J.~V\'egh, ``{Which scaling rule applies to Artificial Neural Networks},'' in
	\emph{{
			Computational Intelligence (CSCE) The 22nd Int'l Conf on Artificial
			Intelligence (ICAI'20)}}.\hskip 1em plus 0.5em minus 0.4em\relax IEEE, 2020,
	pp. Accepted ICA2246, in print; in review in Neurocomputing. [Online].
	Available: \url{http://arxiv.org/abs/2005.08942}
	\BIBentrySTDinterwordspacing
	
	\bibitem{GameOverYelick:2011}
	\BIBentryALTinterwordspacing
	{US National Research Council}, ``{The Future of Computing Performance: Game
		Over or Next Level?}'' 2011. [Online]. Available:
	\url{http://science.energy.gov/~/media/ascr/ascac/pdf/meetings/mar11/Yelick.pdf}
	\BIBentrySTDinterwordspacing
	
	\bibitem{LimitsOfLimits2014}
	I.~Markov, ``{Limits on fundamental limits to computation},'' \emph{Nature},
	vol. 512(7513), pp. 147--154, 2014.
	
	\bibitem{ScalingParallel:1993}
	J.~P. Singh, J.~L. Hennessy, and A.~Gupta, ``Scaling parallel programs for
	multiprocessors: Methodology and examples,'' \emph{Computer}, vol.~26, no.~7,
	pp. 42--50, Jul. 1993.
	
	\bibitem{VeghReevaluate:2020}
	\BIBentryALTinterwordspacing
	J.~V\'egh, ``{Re-evaluating scaling methods for distributed parallel
		systems},'' \emph{IEEE Transactions on Distributed and Parallel Computing},
	vol.~??, p. Refused, 2020. [Online]. Available:
	\url{https://arxiv.org/abs/2002.08316}
	\BIBentrySTDinterwordspacing
	
	\bibitem{Tsafrir:2007}
	D.~Tsafrir, ``The context-switch overhead inflicted by hardware interrupts (and
	the enigma of do-nothing loops),'' in \emph{Proceedings of the 2007 Workshop
		on Experimental Computer Science}, ser. ExpCS '07.\hskip 1em plus 0.5em minus
	0.4em\relax New York, NY, USA: ACM, 2007, pp. 3--3.
	
	\bibitem{armContextSwitching:2007}
	\BIBentryALTinterwordspacing
	F.~M. David, J.~C. Carlyle, and R.~H. Campbell, ``{Context Switch Overheads for
		Linux on ARM Platforms},'' in \emph{Proceedings of the 2007 Workshop on
		Experimental Computer Science}, ser. ExpCS '07.\hskip 1em plus 0.5em minus
	0.4em\relax New York, NY, USA: ACM, 2007. [Online]. Available:
	\url{http://doi.acm.org/10.1145/1281700.1281703}
	\BIBentrySTDinterwordspacing
	
	\bibitem{Gustafson:1988}
	J.~L. Gustafson, ``{Reevaluating Amdahl's Law},'' \emph{Commun. ACM}, vol.~31,
	no.~5, pp. 532--533, May 1988.
	
	\bibitem{VeghMissingSecondDraft:2020}
	\BIBentryALTinterwordspacing
	J.~V\'egh, ``{von Neumann's missing "Second Draft": what it should contain},''
	in \emph{{The 2020 International Conference on Computational Science and
			Computational Intelligence; CSCI'20: December 16-18, 2020, Las Vegas, USA,
			paper CSCI2019}}.\hskip 1em plus 0.5em minus 0.4em\relax IEEE, 2020, p. paper
	CSCI2019. [Online]. Available: \url{https://arxiv.org/}
	\BIBentrySTDinterwordspacing
	
	\bibitem{VerificationBrainScaleS:2020}
	A.~Grubl, S.~Billaudelle, B.~Cramer, V.~Karasenko, and J.~Schemmel,
	``{Verification and Design Methods for the BrainScaleS Neuromorphic Hardware
		System},'' \emph{Journal of Signal Processing Systems}, vol.~92, pp.
	1277--1292, 2020.
	
	\bibitem{Top500:2016}
	{TOP500.org}, ``The top 500 supercomputers,'' \url{https://www.top500.org/},
	2019.
	
	\bibitem{AIcoreProgressStalled:2020}
	M.~Hutson, ``{Core progress in AI has stalled in some fields},''
	\emph{Science}, vol. 368, p. 6494/927, 2020.
	
	\bibitem{NeuralNetworkPerformance:2018}
	S.~J. van Albada, A.~G. Rowley, J.~Senk, M.~Hopkins, M.~Schmidt, A.~B. Stokes,
	D.~R. Lester, M.~Diesmann, and S.~B. Furber, ``{Performance Comparison of the
		Digital Neuromorphic Hardware SpiNNaker and the Neural Network Simulation
		Software NEST for a Full-Scale Cortical Microcircuit Model},''
	\emph{Frontiers in Neuroscience}, vol.~12, p. 291, 2018.
	
	\bibitem{DeepNeuralNetworkTraining:2016}
	\BIBentryALTinterwordspacing
	J.~Keuper and F.-J. Preundt, ``{Distributed Training of Deep Neural Networks:
		Theoretical and Practical Limits of Parallel Scalability},'' in \emph{2nd
		Workshop on Machine Learning in HPC Environments (MLHPC)}.\hskip 1em plus
	0.5em minus 0.4em\relax IEEE, 2016, pp. 1469--1476. [Online]. Available:
	\url{https://www.researchgate.net/publication/308457837}
	\BIBentrySTDinterwordspacing
	
	\bibitem{VeghBiologySpatioTemporal:2020}
	\BIBentryALTinterwordspacing
	{J. V\'egh and \'Ad\'am J. Berki}, ``{On the spatiotemporal behavior in
		biology-mimicking computing systems},'' \emph{Brain Informatics}, p. in
	review, 2020. [Online]. Available: \url{https://arxiv.org/abs/2009.08841}
	\BIBentrySTDinterwordspacing
	
	\bibitem{VeghTemporal:2020}
	\BIBentryALTinterwordspacing
	J.~V\'egh, ``{Introducing Temporal Behavior to Computing Science},'' in
	\emph{{2020 CSCE, Fundamentals of Computing Science}}.\hskip 1em plus 0.5em
	minus 0.4em\relax IEEE, 2020, pp. Accepted FCS2930, in print. [Online].
	Available: \url{https://arxiv.org/abs/2006.01128}
	\BIBentrySTDinterwordspacing
	
	\bibitem{BackusNeumannProgrammingStyle}
	J.~Backus, ``{Can Programming Languages Be liberated from the von Neumann
		Style? A Functional Style and its Algebra of Programs},''
	\emph{Communications of the ACM}, vol.~21, pp. 613–--641, 1978.
	
	\bibitem{ReconfigurableAdaptive2016}
	L.~de~Macedo~Mourelle, N.~Nedjah, and F.~G. Pessanha, \emph{{Reconfigurable and
			Adaptive Computing: Theory and Applications}}.\hskip 1em plus 0.5em minus
	0.4em\relax CRC press, 2016, ch. 5: Interprocess Communication via Crossbar
	for Shared Memory Systems-on-chip.
	
	\bibitem{CommunicationCollapse:2018}
	S.~Moradi and R.~Manohar, ``{The impact of on-chip communication on memory
		technologies for neuromorphic systems},'' \emph{Journal of Physics D: Applied
		Physics}, vol.~52, no.~1, p. 014003, oct 2018.
	
	\bibitem{SpiNNaker:2013}
	S.~B. Furber, D.~R. Lester, L.~A. Plana, J.~D. Garside, E.~Painkras, S.~Temple,
	and A.~D. Brown, ``{Overview of the SpiNNaker System Architecture},''
	\emph{{IEEE Transactions on Computers}}, vol.~62, no.~12, pp. 2454--2467,
	2013.
	
	\bibitem{GordonBellPrize:2017}
	\BIBentryALTinterwordspacing
	G.~Bell, D.~H. Bailey, J.~Dongarra, A.~H. Karp, and K.~Walsh, ``{A look back on
		30 years of the Gordon Bell Prize},'' \emph{The International Journal of High
		Performance Computing Applications}, vol.~31, no.~6, p. 469–484, 2017.
	[Online]. Available: \url{https://doi.org/10.1177/1094342017738610}
	\BIBentrySTDinterwordspacing
	
	\bibitem{InherentSequentiality:2012}
	F.~Ellen, D.~Hendler, and N.~Shavit, ``{On the Inherent Sequentiality of
		Concurrent Objects},'' \emph{SIAM J. Comput.}, vol.~43, no.~3, p. 519–536,
	2012.
	
	\bibitem{PerformanceCounter2013}
	V.~Weaver, D.~Terpstra, and S.~Moore, ``Non-determinism and overcount on modern
	hardware performance counter implementations,'' in \emph{Performance Analysis
		of Systems and Software (ISPASS), 2013 IEEE International Symposium on},
	April 2013, pp. 215--224.
	
	\bibitem{Vegh:2017:AlphaEff}
	J.~V\'egh and P.~Moln\'ar, ``{How to measure perfectness of parallelization in
		hardware/software systems},'' in \emph{18th Internat. Carpathian Control
		Conf.~ICCC}, 2017, pp. 394--399.
	
	\bibitem{PricePerformanceClouds:2017}
	E.~Wustenhoff and T.~S.~E. Ng, ``{Cloud Computing Benchmark},''
	\url{https://www.burstorm.com/price-performance-benchmark/1st-Continuous-Cloud-Price-Performance-Benchmarking.pdf},
	2017.
	
\end{thebibliography}
\end{document}